\documentclass[a4paper]{jpconf}
\usepackage{graphicx}
\begin{document}
\title{Chiral Skyrmionic matter\\
in non-centrosymmetric magnets}

\author{Ulrich K. R\"o\ss ler, Andrei A. Leonov,   Alexei N.\ Bogdanov}

\address{IFW Dresden, Postfach 270116, D-01171 Dresden, Germany}

\ead{u.roessler@ifw-dresden.de}

\begin{abstract}
Axisymmetric  magnetic strings with a fixed sense
of rotation and nanometer sizes (chiral magnetic 
\textit{vortices} or \textit{Skyrmions})
have been predicted to exist in a large group
of non-centrosymmetric crystals
more than two decades ago.
Recently these extraordinary magnetic states 
have been directly observed in thin layers of
cubic helimagnet (Fe,Co)Si.
In this report we apply our earlier theoretical findings 
to review main properties of chiral Skyrmions, 
to elucidate their physical nature, and to analyse 
these recent experimental results 
on magnetic-field-driven evolution of 
Skyrmions and helicoids in chiral helimagnets.
\end{abstract}

\section{Introduction}
In non-centrosymmetric magnetic systems, 
antisymmetric \textit{Dzyaloshinskii-Moriya} (DM) exchange 
causes particular magnetic couplings 
and inhomogeneous states \cite{Dz64}.
These \textit{chiral} interactions  stabilize 
two- and three- dimensional localized structures 
\cite{JETP89,Nature06,JPC10}.
Phenomenologically the inhomogeneous 
Dzyaloshinskii energy contributions \cite{Dz64,PRB02}
are described by Lifshitz invariants,
antisymmetric differential forms 
linear in first spatial derivatives
of the magnetization  $\mathbf{M}$ 
\begin{eqnarray}
\Lambda_{ij}^{(k)} = M_i \frac{\partial M_j}{\partial x_k}
-M_j\frac{\partial M_i}{\partial x_k}
\label{Lifshitz}
\end{eqnarray}
where $x_k$ are Cartesian components of the
spatial variable $\mathbf{r}$.
Solutions for chiral magnetic {\em Skyrmions} 
as static states localized in two dimensions 
have been derived first in 1989 \cite{JETP89}. 
It was shown that these topological 
solitonic field configurations exist in 
magnetic systems for all non-centrosymmetric
crystallographic classes that 
allow Lifshitz invariants in their magnetic 
free energy \cite{JETP89}.
The unique character of these textures stems from the 
general instability of multidimensional solitonic states 
in field theories \cite{Nature06,Derrick64}.
Nonlinear continuum models for condensed
matter systems
do not contain solutions for
static and smooth multidimensional localized states.
Such states appear only as dynamic excitations, 
while static configurations are generally unstable 
and collapse spontaneously into topological singularities. 

For a long time investigations of chiral Skyrmions
have been restricted to theoretical studies
\cite{Nature06,JMMM94,PRB10,confinement10,PRL01}.
The solutions for two-dimensional localized 
and bound states (isolated Skyrmions and Skyrmion lattices)
have been studied in non-centrosymmetric
ferromagnets \cite{JMMM94,PRB10} and antiferromagnets \cite{PRB02},
in cubic helimagnets \cite{Nature06,PRB10,confinement10},
and in confined centrosymmetric
magnetic systems with surface/interface-induced chiral interactions 
(e.g. nanolayers of magnetic metals) \cite{PRL01}.
In \cite{Nature06}, we formulated 
the idea of Skyrmionic matter in
non-centrosymmetric crystals
predicting the existence of mesophases,
composed of Skyrmions as 'molecular units', 
similar to vortex matter 
in type-II superconductors \cite{Blatter94}.
Various effects observed in MnSi and other 
cubic helimagnets with B20 structure \cite{Lebech89} 
indicate multidimensionally modulated magnetic states 
conforming with our theoretical predictions of Skyrmions
and their properties \cite{Nature06,JMMM94}.
The theoretical ideas of Skyrmions in chiral magnets  have 
triggered various experimental efforts to find evidence 
for these twisted textures \cite{Lamago06,Muhlbauer09}.
These experiments collected an impressive range
of data that suggest complex magnetic order phenomena. 
But, mainly using diffraction or indirect evidence 
by transport measurements, these experimental results 
remained essentially inconclusive and have been contested, 
see, e.g., \cite{Maleyev10}. 
Moreover, the interpretation of the experimental
data has been based on approximate solutions \cite{Muhlbauer09,Binz06} 
to the Dzyaloshinskii model by using variational approaches 
in terms of a {\em mode instability}.
The corresponding results do not describe the properties of Skyrmions 
and the phase transition behavior of chiral magnets, 
which is governed by the {\em nucleation} of 
a localized mesoscale entity \cite{Dz64,JMMM94,DeGennes75}.
Our theoretical developments \cite{PRB10,confinement10}
show that multiply modulated chiral states 
in non-centrosymmetric magnets
are composed of localized solitonic states 
with particle-like behavior. 
The thermodynamic stability of condensed Skyrmionic phases 
has been shown \cite{Nature06,PRB10,confinement10}
for the standard and modified Dzyaloshinskii models devised 
for cubic helimagnets \cite{Dz64,Bak80}.
Meanwhile, experimental efforts culminated in
the direct microscopic observations of chiral Skyrmions
in thin layers of (Co,Fe)Si \cite{Yu10}.
This break-through is the first clear experimental 
evidence for the existence of Skyrmions 
as \textit{axisymmetric chiral localized states} that 
are stabilized by a complex interplay of 
nonlinear and chiral effects, as predicted earlier 
\cite{JETP89,Nature06}.

Here, we address the problem of multidimensional
solitonic states in nonlinear systems lacking inversion symmetry.
This emerging field of nonlinear physics 
is based on solutions of nonlinear 
partial differential equations \cite{Nature06,JMMM94,confinement10},
and intricate mathematical methods of micromagnetics \cite{Hubert98},
and physics of solitons \cite{Novikov84}.
Concentrating on the physical side of the problem 
rather than on mathematical details we give 
an elementary introduction into the properties 
of chiral \textit{Skyrmions} in magnetism.

\section{Dzyaloshinskii theory for cubic helimagnets}

\subsection{Phenomenological energy and equations}

Within the phenomenological theory 
introduced by Dzyaloshinskii \cite{Dz64} 
the magnetic energy density of
a  cubic non-centrosymmetric ferromagnet
with the magnetization $\mathbf{M}$
can be written as \cite{Dz64,Bak80}

\begin{equation}
w=\underbrace{A  \left(\mathrm{grad}\mathbf{M} \right)^2
-D\,\mathbf{M}\cdot \mathrm{rot}\mathbf{M}
-\mathbf{M}\cdot\mathbf{H}}_{w_0}
-\sum_{i=1}^{3}\left[B  \left(\partial M_i / \partial x_i \right)^2
+K_c M_i^4 \right] - f(M)
\label{density}
\end{equation}
including the exchange 
stiffness with constant $A$.
The inhomogeneous DM exchange with constant $D$
is a combination of Lifshitz
invariants (\ref{Lifshitz}),
$w_D = D\,(\Lambda_{yx}^{(z)}+\Lambda_{xz}^{(y)}+\Lambda_{zy}^{(x)})$
=$D\,\mathbf{M}\cdot \mathrm{rot}\mathbf{M}$.
These isotropic interactions
together with the Zeeman energy
are the \textit{essential} 
couplings to stabilize
chiral  modulations ($w_0$).
The sum in Eq.~(\ref{density})
includes exchange ($K$) and
cubic ($K_c$) magnetic anisotropy
contributions; $f(M)$ comprises
magnetic interactions imposed
by variation of the magnetization
modulus $M \equiv |\mathbf{M}|$.
In a broad temperature range
the magnetization vector practically does not
change its length, and nonuniform magnetic states
display only rotation of $\mathbf{M}$.
%
%
The low temperature properties, thus,
can be described by
the model (\ref{density}) 
with a fixed magnetization
modulus $M$ = const.
%
%

\subsection{One-dimensional chiral modulations: cones and helicoids}

\begin{figure*}
\includegraphics[width=16cm]{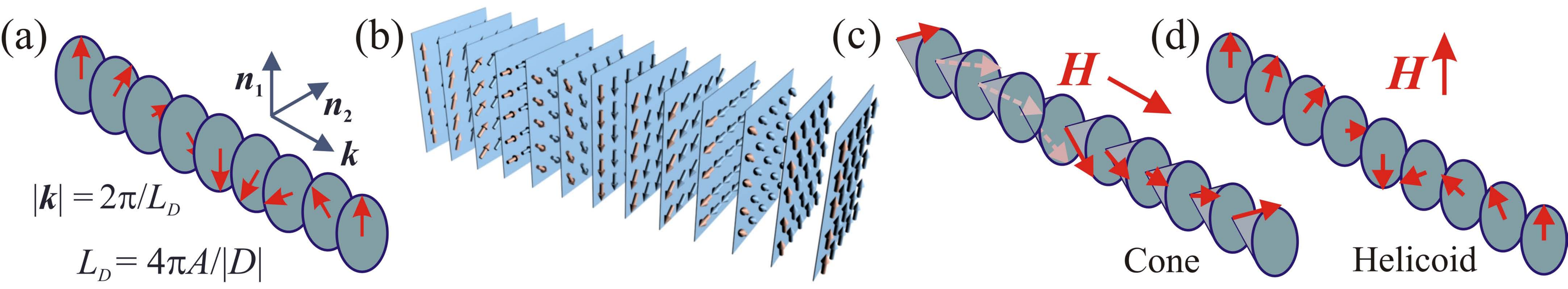}
\caption{
\label{helices} 
One-dimensional chiral modulations in cubic helimagnets.
In a helical "array" (a) the magnetization
rotates in the plane spanned by the orthogonal
unity vectors 
$\mathbf{n}_1$ and $\mathbf{n}_2$ 
and the rotation sense is determined
by the sign of Dzyaloshinskii constant $D$.
Chiral \textit{helices} (b) are composed 
of arrays (a) with the same propagation
direction.
Under the influence of the applied field
the helix (b) transforms
into longitudinal distorted \textit{cones} (c) 
or into transversally distorted 
\textit{helicoids}(d). 
}
\end{figure*}

In non-centrosymmetric magnets,
chiral couplings of type (\ref{Lifshitz}) favour
rotation of the magnetization vector.
They destabilize the homogeneous magnetic structure
and induce long-range modulations of the magnetization \cite{Dz64}.
At zero field helical modulations
\begin{equation}
\mathbf{M}= M \left[ \mathbf{n}_1 \cos 
\left(\mathbf{k} \cdot \mathbf{r} \right)
+ \mathbf{n}_2 \sin  \left(\mathbf{k}
\cdot \mathbf{r} \right) \right], 
\quad  |\mathbf{k}| = 2 \pi/L_D, 
\quad  L_D = 4 \pi A/|D|
\label{helix0}
\end{equation}
with period $L_D$ and wave vector
$\mathbf{k}$  correspond to the absolute
minimum of the  isotropic functional $w_0$
(Fig. \ref{helices} a, b).
In Eq. (\ref{helix0}) $\mathbf{n}_1$, $\mathbf{n}_2$ 
are orthogonal unit vectors 
in the plane of the magnetization rotation (Fig. 1). 
The modulations (\ref{helix0})
have a fixed rotation sense determined by the sign
of constant $D$ and are continuously degenerate 
with respect to propagation directions in the space.

An applied magnetic field
lifts the degeneracy of the helices (\ref{helix0})
and stabilizes a {\em cone} solution with the propagation 
direction along the magnetic field 
\begin{equation}
\cos \theta = |\mathbf{H}|/H_D,  \quad \psi = 2 \pi z/L_D,
\quad H_D = D^2 M/(2A),
\label{cone1}
\end{equation}
where $\mathbf{M}=M(\sin\theta\cos\psi;
\sin\theta\sin\psi;\cos\theta)$ is written
in spherical coordinates.
In such a helix the magnetization component along
the applied field has a fixed value 
$M_{\bot} = M \cos \theta = M(H/H_D)$, and
the magnetization vector $\mathbf{M}$ rotates
within a cone surface.
The critical value $H_D$ marks the saturation
field of the cone phase.
The helix periods ($L_D$) and critical fields ($H_D$)
for some non-centrosymmetric
cubic ferromagnets (\textit{helimagnets}) are
presented in Table~\ref{table1}.

\begin{table}[h]
\caption{
\label{table1}
N{\'e}el temperatures ($T_N$), helix periods ($L_D$), and saturation fields ($H_D$) \\
for some cubic helimagnets, data from Refs. \cite{Lebech89}.}
\begin{center}
\begin{tabular}{llllllllll}
\br
Compound \quad \quad & MnSi \quad \quad & FeGe & Fe$_{0.3}$Co$_{0.7}$Si \quad \quad
 & Fe$_{0.5}$Co$_{0.5}$Si \quad \quad & Fe$_{0.8}$Co$_{0.2}$Si \\
 \mr
$T_N$ [K] & 29.5& 278.7& 8.8 & 43.5& 32.2\\
$L_D$ [nm] & 18.0 & 68.3- 70.0 & 230&  90.0 & 29.5 \\
$H_D$ [T] & 0.62 & 0.2 & (6.0 $\pm$ 1.5)$\cdot 10^{-3}$ &  (4.0 $\pm$ 0.5)$\cdot 10^{-2}$  & 0.18 \\
\br
\end{tabular}
\end{center}
\end{table}
%
In real non-centrosymmetric magnets anisotropic
forces  fix the propagation of chiral modulations
along certain \textit{easy axes} directions
imposed by the crystal symmetry.
A magnetic field perpendicular
to the propagation direction violates
the uniform rotation of the magnetization
($\psi \propto z$, Eq.~(\ref{helix0})).
In a sufficiently high magnetic field 
$H = H_H$  distorted \textit{helicoids}  
infinitely expand their period 
into a system of isolated 2$\pi$- domain 
walls separating domains with the magnetization
along the applied field 
(Figs. \ref{lattice} (d), \ref{mcurves} (a)) 
\cite{Dz64,JMMM94}.

\section{Chiral flux-lines: the building blocks for Skyrmionic textures}

\subsection{Chiral Skyrmions in cubic helimagnets}

\begin{figure*}
\includegraphics[width=16cm]{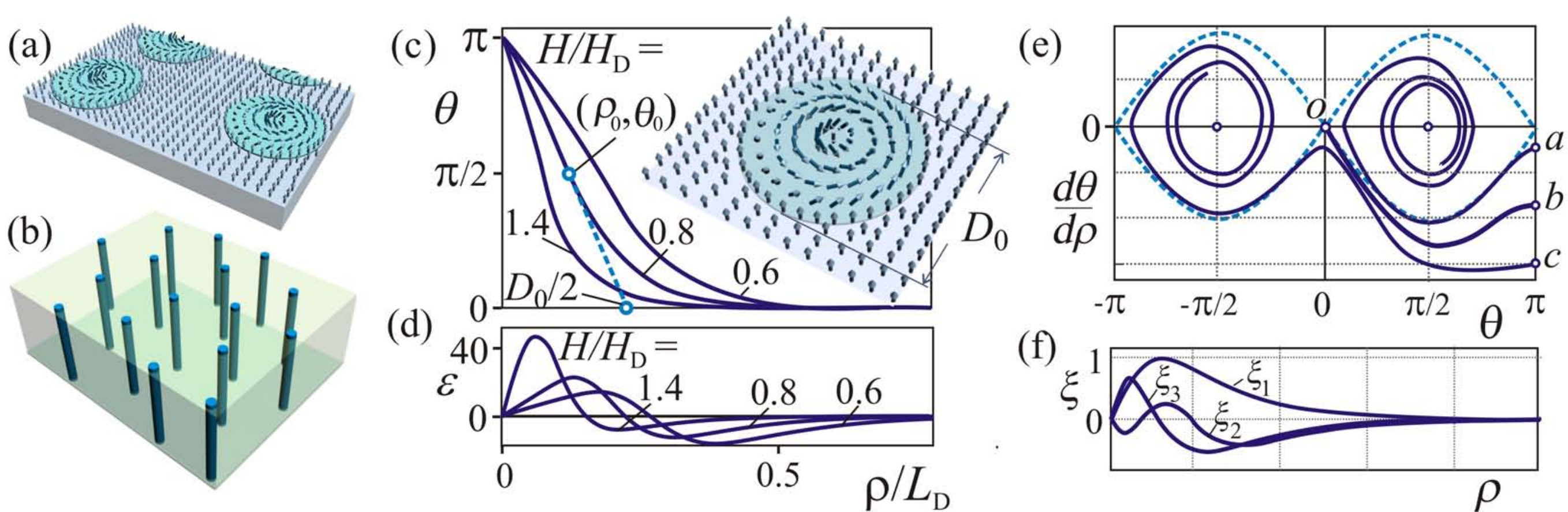}
\caption{Isolated skyrmions in  a nanolayer (a)
and a bulk sample (b) of a cubic helimagnet.
The magnetization profiles $ \theta (\rho)$ (c)
and corresponding  energy densities  
$\varepsilon(\rho)$ (d) for
typical solutions of Eq. \ref{eq}.
A cross-section through an isolated skyrmion
shows axisymmetric distribution of the
magnetization (shaded area indicates the core
with the diameter $D_0$ (Inset (c)).
Phase trajectories of skyrmions correspond
to separatrix curves in the phase space 
(curve $b-o$)(e).
Typical eigen functions for
radial excitations (f).
}
\label{Skyrmion}
\end{figure*}

The equations minimizing energy (\ref{density})
include axisymmetric localized
solutions $\psi = \varphi + \pi/2$,
$\theta = \theta(\rho)$, $M$ = const
(Fig. \ref{Skyrmion}, (a), (b))
where the spatial variable
$\textit{\textbf{r}}=
(\rho\cos\varphi;\rho\sin\varphi;z)$
is written in cylindrical coordinates and
the magnetization in spherical coordinates
(\ref{cone1}).
The Euler equation for $\theta(\rho)$
to determine the equilibrium structure of isolated Skyrmions
 \cite{JETP89,JMMM94} is
\begin{eqnarray}
\label{eq}
A \left( \frac{d^2 \theta}{d \rho ^2} 
+ \frac{1}{\rho}\frac{d \theta}{d \rho} \right.
\left. - \frac{1}{\rho ^2}\sin \theta \cos \theta  \right)
-\frac{H}{2M}\sin \theta 
-\frac{D}{\rho} \sin^2 \theta=0
\end{eqnarray}
with boundary conditions 
$\theta(0)=\pi,\, \theta(\infty)=0$.
\begin{figure*}
\includegraphics[width=16cm]{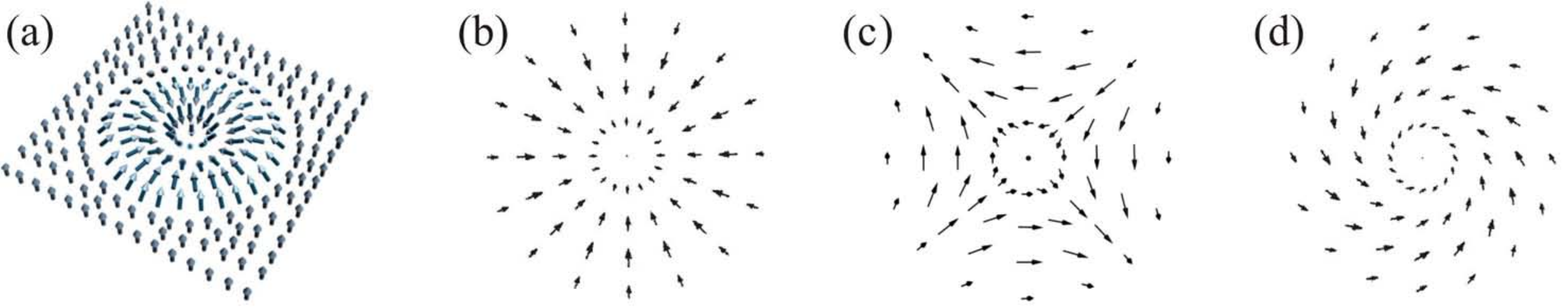}
\caption{Magnetization arrangement in Skyrmions 
of uniaxial ferromagnets with $n$mm (C$_{nv}$) (a,b), 
$\bar{4}$2m (D$_{2d}$)(c), and $n$ (C$_n$) (d) symmetries.
}
\label{uniaxial2}
\end{figure*}
In \textit{non-centrosymmetric uniaxial magnets}  functionals $w_D$
have anisotropic forms depending on
crystallographic class and may include
several terms \cite{JETP89}.
For example, in ferromagnets belonging to
$C_n$ crystallographic classes ($n = 3,4,6$) 
the chiral energy $w_D$ includes three Dzyaloshinskii constants
$D_i$ \cite{JETP89},
\(w_D=D_1\,(\Lambda_{xz}^{(x)}+\Lambda_{yz}^{(y)}) +
D_2\,(\Lambda_{xz}^{(y)}-\Lambda_{xz}^{(y)}) +
D_3\,\Lambda_{xy}^{(z)}\).
For all non-centrosymmetric ferromagnets
the equation for isolated Skyrmions $\theta (\rho)$ 
have the same functional form as Eq. (\ref{eq}),
however, $\psi(\varphi)$ have different solutions
(Fig. \ref{uniaxial2})
\begin{eqnarray}
\label{uniskyrm1}
\psi = \varphi,  \   (nmm), \quad
\psi = -\varphi+ \pi/2, \ (\bar{4}2m),
\quad
\psi = \varphi + \pi/2,  \ (n22),
 \quad
\psi = \varphi + \gamma,  \ (n)
\end{eqnarray}
where $\gamma = \arctan (D_1/D_2)$.
For a complete list of $w_D$ functionals and solutions $\psi (\varphi)$,
see Ref.~\cite{JETP89}).
\subsection{Solutions for Skyrmions}
Typical solutions of Eq.~(\ref{eq}), $\theta (\rho)$,
consist of arrow-like cores
($ \pi - \theta \propto \rho$ for  $\rho \leq L_D$)
and  exponential "tails",
$ \theta \propto  \exp{(-\rho)}$ for
$\rho \gg L_D$ \cite{JETP89,JMMM94}
(Fig. \ref{Skyrmion}).
In  \textit{phase space}
$(\theta, d \theta / d \rho)$ 
these localized solutions correspond 
to \textit{separatrix} trajectories
(e.g. to the curve with initial derivative
$(d \theta / d \rho)_0 = b$ in Fig. \ref{Skyrmion}(e)). 
A Skyrmion core diameter $D_0$ can be defined 
as two times the value of $R_0$, which is the coordinate 
of the point where the tangent at the inflection 
point ($\rho_0, \theta_0$) intersects the $\rho$-axis
(Fig.~\ref{Skyrmion}(d)), 
in analogy to definitions for domain wall width \cite{Hubert98}.
Skyrmion energy per unit length and diameter are
\begin{eqnarray}
\label{energySkyrmion}
E = 2\pi \int_{0}^{\infty} \varepsilon (\theta, \rho) d \rho,
\quad \quad
D_0= 2L_0[\rho_0 - \theta_0 (d \theta / d \rho)^{-1}_{\rho = \rho_0}]
\end{eqnarray}
where $\varepsilon (\theta, \rho)$ is the "linear"  energy density
\cite{JMMM94}. 
The energy density distributions $\varepsilon ( \rho)$ 
(Fig.~\ref{Skyrmion}(f)) reveal two distinct
regions: positive energy "bags" are concentrated
in the Skyrmion center and are surrounded by
extended areas with negative energy density,
where the DM exchange dominates.
The radial stability of Skyrmion solutions has been
proved by solving the corresponding spectral
problem \cite{JMMM94}, which also yields
excitation modes of Skyrmion cores, as 
in Fig. \ref{Skyrmion}(f), 

\subsection{Analytical results for the linear ansatz}
For localized solutions $\theta (\rho)$ a linear trial
function has proved to be a suitable approximation \cite{JETP89}.
With ansatz \(\theta = \pi (1 - \rho/R) \ \  ( 0 < \rho < R), \quad
\theta = 0  \  \  (\rho > R)\)
the Skyrmion energy (\ref{energySkyrmion}) is
reduced to a quadratic potential
\begin{eqnarray}
\label{ansatz2}
E (R)  = E_0 + \frac{\alpha H}{M} R^2  - \frac{\pi}{2} |D| R,
\quad  \quad 
 R_{\mathrm{min}} = 0.42 L_D  H_D/H,
\quad
E_{\mathrm{min}} = E_0 - 4.15 H_D/H
\end{eqnarray}
where
$E_0 = [\pi^2 + \mathrm{Cin}\,2 \pi]A/2 = 6.154\,A$, 
$\alpha = (1-  4/\pi^2)/2 =0.297$,
and the parabola vertex 
$(R_{\mathrm{min}}, E_{\mathrm{min}})$ 
determines the minimum of energy (\ref{ansatz2}).
This simplified model offers 
an important insight into the mechanisms underlying 
the formation of chiral Skyrmions.
The exchange energy $E_0$ does not depend
on the Skyrmion size and contributes a 
positive energy associated with its distorted core.
The equilibium Skyrmion size 
arises as a result of the competition between
chiral and Zeeman energies:
$R_{\mathrm{min}} \propto |D|/H$,
and becomes zero in centrosymmetric systems ($D =0$).
Isolated solutions of Eq.~(\ref{eq}) with 
positive energy exist at high applied magnetic
fields. With decreasing fields 
equilibrium energy $E_{\mathrm{min}})$ for Eq.~(\ref{ansatz2}) 
decreases and becomes \textit{negative}
at a critical value $H_S$.
Below $H_S$ the Skyrmions condense into a lattice
at the exact value $H_S$=0.80132 (Table \ref{table2})) \cite{JMMM94}.

\section{Skyrmion lattices: where 
does "double-twist" become beneficial?}
\subsection{Solutions for bound Skyrmions}
\begin{figure*}
\includegraphics[width=16cm]{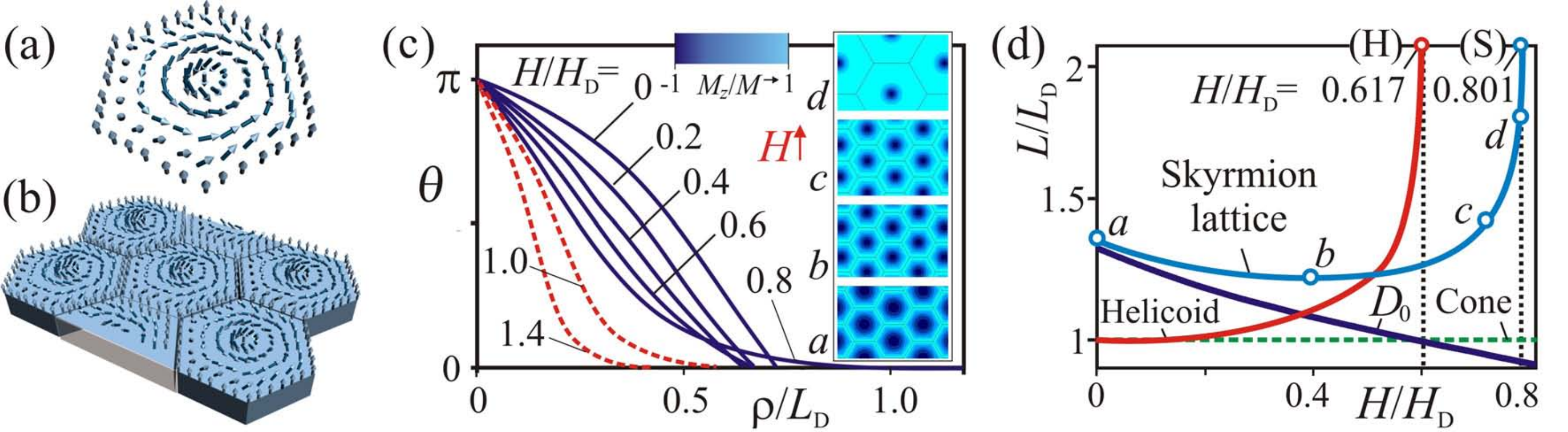}
\caption{Skyrmion lattices:
(a) unit cells have axisymmetric
magnetization structure near the center;
(b) a thin layer with a hexagonal lattice.
The magnetization
profiles  $\theta(\rho/L_D)$ of a
Skyrmion cell (solid blue lines)
and of an isolated Skyrmion
(dashed red lines) for different values
of applied magnetic field (c). 
Equilibrium sizes of the cell core ($D_0$)
and lattice period compared to helicoid periods (d).
Inset shows the transformation
of the hexagonal lattice into a
set of isolated Skyrmions in 
increasing magnetic field.
Snapshots correspond to the
field values indicated 
in the diagram for period $L/L_D$ vs. applied field $(H/H_D)$ 
by hollow circles (d).
}
\label{lattice}
\end{figure*}

The equilibrium parameters of a 
Skyrmion lattice are derived from a system 
of differential equations for 
$\theta (x,y)$, $\psi (x,y)$ minimizing 
the system energy. 
For Skyrmions at low temperatures
the \textit{circular cell} approximation
provides a very elegant and reliable 
approximation \cite{JMMM94}.
In this method the lattice cell
is replaced by a circle of equal area.
The equilibrium parameters of Skyrmion lattices
are derived by integration of Eq.~(\ref{eq})
with boundary conditions $\theta(0) = \pi$,
$\theta(R) = 0$ and a subsequent minimization 
of the lattice energy density
$W_{S} =(2/R^2)\int^{R}_{0} w(\rho)\rho d\rho$
with respect to the cell radius $R$ \cite{JMMM94}.
The solutions for magnetization profiles $\theta( \rho)$ (c) and
the lattice periods (d) are plotted in Fig. \ref{lattice}.
The reduced perpendicular magnetization of 
a Skyrmion lattice averaged over the Skyrmion
cell $m_{\small{S}} = (2/R^2)\int_0^{R} \cos(\theta) \rho d \rho$
is plotted together with the magnetization of the helicoid,
$m_{\small{H}} $  (Fig.~\ref{mcurves}).
Contrary to the helicoid the Skyrmion lattice has a finite
magnetization at zero field, $m_{\small{S}}  (0)$ = 0.124.

\begin{table}[h]
\caption{
\label{table2}
Critical fields and characteristic parameters of the hexagonal Skyrmion lattice:
$H_1$ transition field between the helicoid and Skyrmion lattice; 
$H_S$ saturation field of the Skyrmion lattice; last column for 
isolated Skyrmions as excitations of the saturated state.
}
\begin{center}
\begin{tabular}{llllll}
\br
&& $H_1$& $H_{S}$& \\
 Reduced magnetic field, $H/H_D \quad \quad $&$0$&$0.216 \quad \quad$&$0.801 \quad \quad$&1.4\\
 \mr
Lattice cell period, $L/L_D$&$1.376 \quad \quad$&1.270& $\infty$ & - \\
Core diameter, $D_0/L_D$&1.362&1.226& 0.920 & 0.461 \\
Averaged magnetization, $m_{\small{S}}$ &0.124& 0.278&1&1  \\
\br
\end{tabular}
\end{center}
\end{table}
\subsection{Skyrmions compete with helicoids. Transition field $H_1$}
\begin{figure}[h]
\includegraphics[width=16 cm]{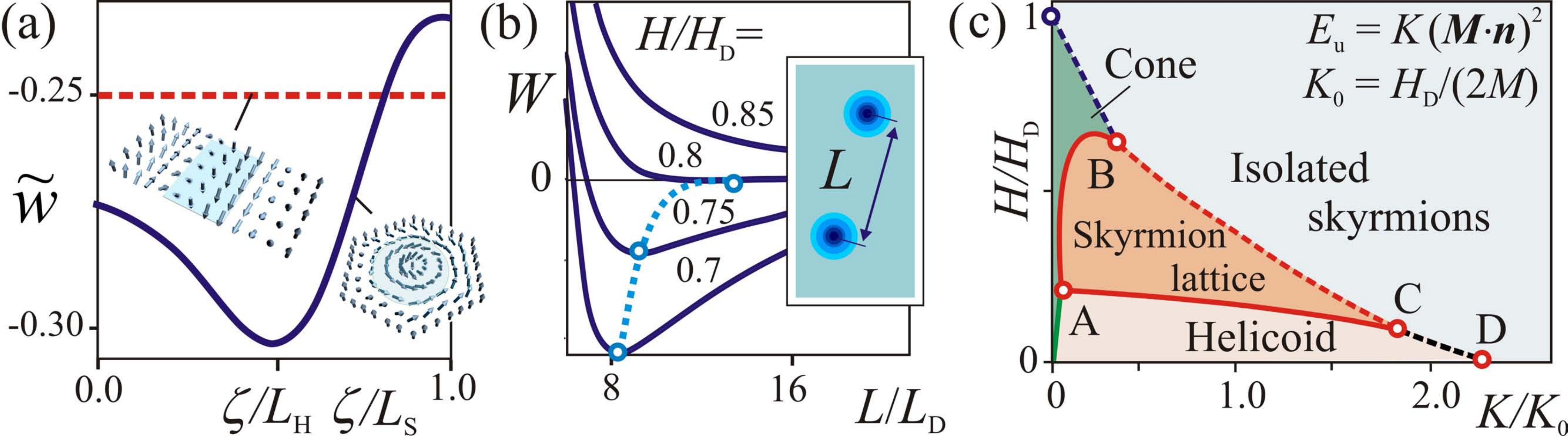}
\caption{\label{uniaxial0}
Local energies $\widetilde{w}(\zeta)$ of
the Skyrmion lattice and helicoid at zero field (a).
Below the critical field $H_S$ the negative
internal energies of Skyrmions overcome the repulsive
Skyrmion-Skyrmion interaction and a dense packed 
(hexagonal) lattice is created (b).
Under the influence of induced uniaxial anisotropies ($E_u$)
the Skyrmion lattice becomes globally stable in 
a broad range of the applied fields
($K_0$ is the critical value for uniaxial
distortions suppressing the cones in zero field) 
(adopted from \cite{PRB10}) (c).
}
\end{figure}

In the Skyrmion lattices rotation of the magnetization in two directions
leads to a larger reduction in the Dzyaloshinskii-Moriya energy than
single-direction rotation in helical phases.
On the other hand, such
\textit{double-twist} modulations increase the exchange energy.
The equilibrium energy of the Skyrmion cell at zero field
$\widetilde{w}_{S}(\zeta )=(2/\zeta^2)\int^{\zeta}_{0} w(\rho)\rho d\rho$
plotted as a function of the distance from the center $\zeta$
(Fig. \ref{uniaxial0} (a)) shows that an energy excess near 
the border outweighs the energy gain 
at the Skyrmion center (for details see \cite{Nature06}). 
At higher magnetic fields, however, the Skyrmion lattice has 
lower energy than the helicoid. The first order transition 
between these two modulated states occurs at $H_1 = 0.2168 H_D$  \cite{JMMM94}.

\subsection{"Receding" Skyrmions. Critical field $H_S$}
Properties of the Skyrmion lattice solutions 
are collected in Figs.~\ref{lattice}, \ref{mcurves}
and in Table \ref{table2}.
With increasing magnetic field, 
a gradual localization of the Skyrmion core $D_0$ 
is accompanied  by the expansion of the lattice period.
The lattice transforms
into the homogeneous state by infinite
expansion of the period at the critical field $H_S = 0.80132 H_D$.
Remarkably, 
the Skyrmion core retains a finite size,
$D_0 (H_S) = 0.920 L_D$ and
the lattice releases a set 
of \textit{repulsive} isolated Skyrmions
at the transition field $H_S$,
owing to their topological stability.
These free Skyrmions can exist 
far above $H_S$. 
On decreasing the field again below $H_S$, 
they can re-condense into a Skyrmion lattice 
(Fig. \ref{uniaxial0} (b)).
A similar type of sublimation and resublimation 
of particle-like textures occurs in
helicoids at the critical field 
$ H =H_H /H_D= \pi^2/8 = 0.6168$:
the period infinitely expands and the helicoid
splits into a set of isolated 2$\pi$ domain
walls or kinks \cite{Dz64,JMMM94}.

This peculiar transformation of chiral modulations
into homogeneous states constitutes 
a \textit{nucleation} type of continuous phase transition,
according to a classification introduced by 
De Gennes \cite{DeGennes75}.
In contrast to the common \textit{instability} 
type of continuous (2nd order) transitions
that is described by the amplitude of a sole 
fundamental mode acting as the order parameter, 
the nucleation transitions retain a full spectrum 
of lattice modes at the transition. 
Magnetic-field-driven transitions of multidomain
states into the homogeneous phase also belong 
to the nucleation type of phase transformations \cite{Hubert98}.

\subsection{Stabilization effects of magnetic anisotropy}
For isotropic model $w_0$ (\ref{density}) the cone
phase (\ref{cone1}) is the global minimum in the whole range 
of the applied fields where the modulated
states exist ($ 0 < H < H_D$). 
The helicoids and Skyrmion lattices can exist 
only as metastable states. 
Under the influence of anisotropies (\ref{density}),
however, the Skyrmion
lattices may gain thermodynamic stability
in a certain interval of applied
fields, for details see \cite{PRB10,confinement10}.
E.g., uniaxial anisotropies can be induced 
in bulk helimagnets by uniaxial stresses or they
arise due to surface effects in thin layers.
These anisotropies are described by 
$ E_u = K (\mathbf{M}\cdot \mathbf{n})^2$,
where $K$ is a constant of uniaxial anisotropy,
$\mathbf{n}$ is a unity vector along the distortion axis).
Such uniaxial distortions
suppress the cone phase in cubic helimagnets 
and establish global stability of the Skyrmion lattice in broad
ranges of the thermodynamic parameters 
(Fig. \ref{uniaxial0} (c)) \cite{PRB10}.

\section{On the observation of  Skyrmionic and helical
textures in Fe$_{0.5}$Co$_{0.5}$Si nanolayers}

\begin{figure*}
\includegraphics[width=16cm]{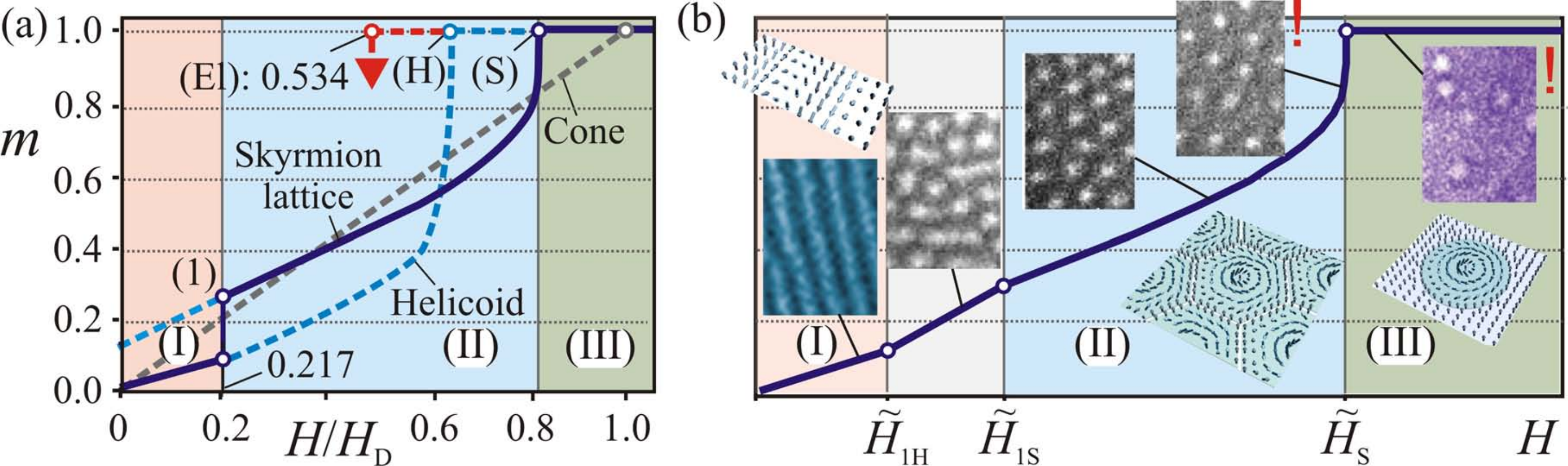}
\caption{The ideal magnetization curves
for a bulk sample (based on results of
\cite{JMMM94}) (a) and for a thin layer (b)
of a cubic helimagnet with suppressed cone phases.
Solid lines indicate the thermodynamically
stable states; dashed lines in Fig. (a), metastable
configurations.
The 1st order transition line at $H_1$ 
in the thin layer
is expanded into a region of multidomain states.
Fragments of experimentally observed images
\cite{Yu10} are in complete agreement
with theoretically calculated magnetization 
curves. The patterns marked by ``(!)''
display \textit{isolated chiral Skyrmions}.
}
\label{mcurves}
\end{figure*}

\vspace{2mm}
Real-space images of Skyrmion states 
in a thin layer of cubic helimagnet Fe$_{0.5}$Co$_{0.5}$Si
have recently been obtained by using Lorentz transmission
electron microscopy \cite{Yu10}. 
This is the first clear experimental
manifestation of \textit{chiral Skyrmion states}.
The first-order transition
of a helicoid into a Skyrmion lattice and  its subsequent transformation
into a system of isolated Skyrmions  observed in bias magnetic
fields (Figs. 1, d-f, 2, 3 (a-d) in \cite{Yu10})
are in excellent agreement with the theoretical
predictions on the behavior of Skyrmions and the field-driven 
transitions into densely packed Skyrmion lattices
according to the magnetic phase diagrams calculated
earlier \cite{JMMM94,Nature06} (Fig.~\ref{mcurves}).

In the experiments, the thickness 20~nm of 
the magnetic layer Fe$_{0.5}$Co$_{0.5}$Si 
is much smaller than the helix period $L_D$ = 90 nm \cite{Yu10}.
But, even for such a small thickness, the conical state  
propagating only for a fraction of a period perpendicularly
through the layer has lower energy than a Skyrmion lattice, 
absent additional effects that stabilize it in applied fields. 
Usually in magnetic nanolayers strong perpendicular 
uniaxial anisotropy arises, either as a result of surface 
effects \cite{Johnson96} or of lattice strains.
Thus, induced anisotropies give a possible explanation 
for the experimental observation of the Skyrmions in
these Fe$_{0.5}$ Co$_{0.5}$Si layers, in accordance with
the phase diagram for cubic helimagnets with uniaxial 
distortions (Fig.~\ref{uniaxial0} (c)) \cite{PRB10}.
Fig.~\ref{mcurves} presents the magnetization curve for a bulk isotropic
helimagnet (a) (based on results of \cite{JMMM94}, Fig. 12)
and the corresponding magnetization curve for a thin layer
involving demagnetization effects \cite{Hubert98} (b). 
Compared to theoretically calculated
values in a bulk material ($H_S$, $H_H$) the
corresponding critical fields in a thin layer
are shifted, and their values can be estimated
as $\widetilde{H}_{S(H)} = H_{S(H)} + 4\pi M$.
Due to demagnetization effects multidomain states
can be stabilized in the vicinity of the transition 
field $H_1$ \cite{JMMM94}. The boundaries of these 
regions with coexisting phases can be estimated as
$\widetilde{H}_{1H}= H_1 + 4\pi M m_{\small{H}}(H_1) $,
$\widetilde{H}_{1S}= H_1 + 4\pi M m_{\small{H}}(H_1) $.
The magnetizations of the competing phases at the transition field
equal  $m_{\small{H}}  (H_1)$ = 0.111 and $m_{\small{S}}  (H_1)$ = 0.278.
The jump of the magnetization during the transition equals
$\Delta M = [ m_{\small{S}} (H_1) - m_{\small{H}} (H_1)] M = 0.167 M$,
i.e., it reaches about 17 \% of the saturation value.  

The magnetization curves in Fig.~\ref{mcurves} are constructed
for ideally soft magnetic material under the condition that only
the equilibrium states are realized in the magnetic sample.
In real materials the formation of the equilibrium 
states  is often hindered (especially during the phase transitions), 
and evolution of metastable states and hysteresis effects play 
an important role in the magnetization processes.
Particularly, the formation of the Skyrmion lattice below $H_S$
can be suppressed. Then isolated Skyrmions exist below this
critical field. At a critical field H$_{El}$
the Skyrmions become unstable with respect to elliptical 
deformations and "strip-out" into isolated 2$\pi$ domain walls.
In a bulk material $H_{El} = 0.534 \, H_D$ 
(indicated in Fig.~\ref{mcurves} with a red arrow). 
In a thin layer, one estimates
$\widetilde{H}_{El} = H_{El} + 4 \pi M$.
As discussed earlier \cite{JETP89,Nature06,JMMM94}
the evolution of chiral Skyrmions in magnetic fields 
has many features in common with that of bubble domains
in perpendicular magnetized films,\cite{Hubert98} and
with Abrikosov vortices in superconductors \cite{Blatter94}.
 
The fragments of images from Ref. \cite{Yu10} (Fig.\ref{mcurves} (b))
reflect in details theoretically
predicted evolution of the chiral modulations
in the applied magnetic field:
the helicoid phase is realized at low fields (region (I));
at higher field this transforms into the Skyrmion lattice
(region (II)) via an intermediate state
($\widetilde{H}_{1H}< H < \widetilde{H}_{1S}$);
finally the Skyrmion lattice by extension of the
period transforms into the homogeneous phase 
where isolated Skyrmions still exist as topologically
stable 2D solitons.

Two patterns indicated in Fig.~\ref{mcurves} (b)
with exclamation mark manifest the main result
of Ref.~\cite{Yu10}:  the first images of 
\textit{static two-dimensional localized states}
aka \textit{chiral Skyrmions}!
In Ref.  \cite{Yu10} this result has been overlooked
and misinterpreted as a coexisting ferromagnetic 
and Skyrmion lattice phases.
As it was expounded in the previous section, the
transition of the Skyrmion lattice into the 
homogeneous state is a \textit{continuous} transition,
but of the particular \textit{nucleation} type.
Such transitions exclude the formation of coexisting states.

The condensed Skyrmion phases in the micrograph of  Ref.~\cite{Yu10}
also appear as heavily distorted densely packed two-dimensional
lattice configurations. This is expected for Skyrmionic matter. 
As these mesophases are composed from elastically
coupled radial strings, dense Skyrmion configurations 
generally do not form ideal crystalline lattices 
but various kinds of partially ordered states, 
e.g. hexatic ordering implying only orientational order
of bonds without positional long-range order, or other 
glassy  arrangement following standard arguments 
put forth for the similar vortex matter in type-II superconductors
\cite{Blatter94}.
The observation derives from the particle-like (or string-like)
nature of Skyrmions and suggests that Skyrmionic mesophases may
display rich phase diagrams. 

\section{Chiral modulations near the ordering temperature}

\subsection{Solutions for high-temperature Skyrmions. 
Confinement temperature $T_p$}

Near the ordering temperature the magnetization
amplitude varies under the influence of the applied
field and temperature. Basic properties of chiral
modulations in this region can be derived by
minimization of isotropic model
$w = w_0 - f(M)$ (\ref{density}) with respect
to \textit{all} components of the magnetization vector
($M, \theta, \psi$).
Within this model  axisymmetric isolated structures
are described by equations  
$\psi (\varphi)$ (\ref{uniskyrm1})
and the solutions of the Euler equations
for $\theta(\rho)$, $M(\rho)$ \cite{Nature06,confinement10} 
\begin{eqnarray}
A \left( \frac{d^2 \theta}{d \rho^2} +  \frac{1}{\rho} \frac{ d \theta}{d \rho}
-\frac{\sin \theta \cos \theta}{\rho^2} \right)
 - D\frac{2}{\rho}\sin ^2 \theta - H\sin \theta  
 + A \frac{2}{M} \frac{d M}{d\rho} \left( \frac{ d \theta}{d \rho} -1 \right) = 0,
\nonumber \\
A \left[\frac{1}{M} \left(\frac{d^2 M}{d \rho^2} +  \frac{1}{\rho} \frac{ d M}{d \rho} \right)
+\left(\frac{ d \theta}{d \rho}\right)^2 + \frac{\sin^2 \theta}{\rho^2} \right]
+D \left[ \frac{ d \theta}{d \rho} +\frac{\sin \theta \cos \theta}{\rho} \right] 
-\frac{H}{M} \cos \theta + \frac{1}{M} \frac{\partial f }{\partial M }= 0
\label{Euler1}
\end{eqnarray}
With $ f(M) = a_1 M^2 + a_2 M^4$ 
($a_1 = J (T_c-T)$, $a_2 > 0$) which is commonly
used to describe magnetization processes in cubic helimagnets at
high temperatures the solutions of Eqs. (\ref{Euler1}) have been
investigated in \cite{Nature06,confinement10}.
With boundary conditions $\theta(0) = \pi$, $\theta(\infty) = 0$,
$M(\infty) = M_0 $, Eqs.~(\ref{Euler1}) describe the structure of isolated Skyrmions.
The magnetization of the homogeneous 
phase $M_0$ is derived from equation  $2a_1 M+ 4 a_2 M^3 - H =0$.
It is convenient to introduce new parameters,
an effective temperature $a = (T_c - T)/T_D$ and
$H_0 =  \left(D^2/2 A\right)^{3/2}/\sqrt{a_2}$ where
"chiral shift", $T_D = D^2/(8 A J)$ is a characteristic
temperature equal to the difference between the ordering temperature
of the helimagnet ($T_N = T_c + T_D$) and the ferromagnet
with $D =0$ ($T_c$).

\begin{figure*}
\includegraphics[width=16cm]{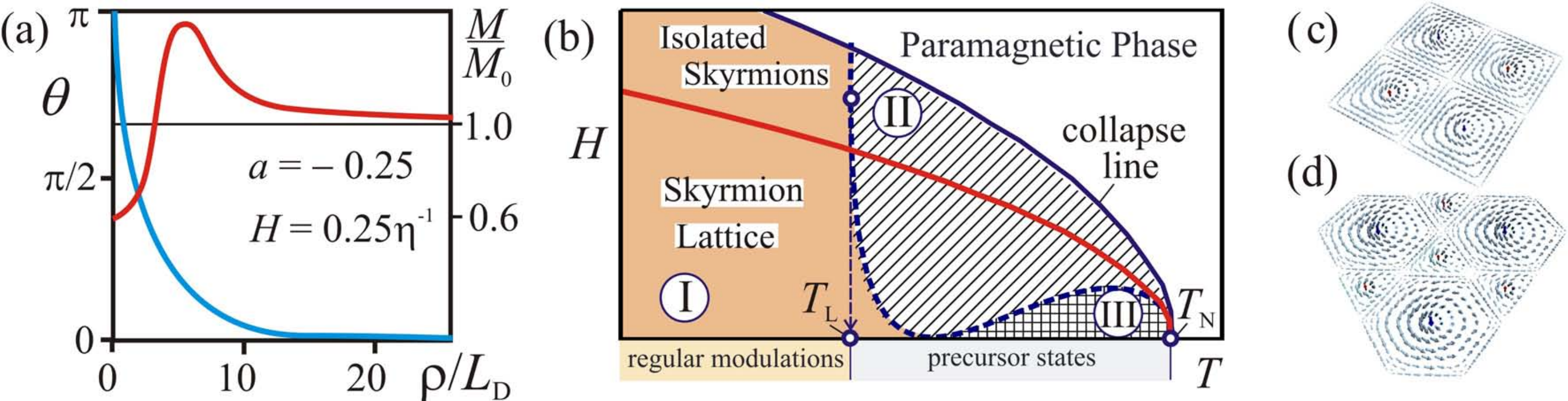}
\caption{Typical solutions 
for magnetization profiles $\theta (\rho)$,
$M(\rho)$ (Eqs. \ref{Euler1}) (a).
Magnetic phase diagram of  an \textit{isotropic}
helimagnet  near the ordering temperature
includes areas with \textit{repulsive} (I),
\textit{attractive} (II) skyrmions, and
\textit{confined} skyrmion states (III) (b).
\textit{Confinement temperature} $T_L$ separates
the main part of phase diagram with regular
chiral modulations ($T < T_L$) from the region
of "precursor state" ($T_L < T < T_N$).
The confined skyrmion textures include square
(\textit{half}-skyrmion) (c) and hexagonal
(d) lattices.
}
\label{conf1}
\end{figure*}

Typical solutions  $\theta (\rho)$, $M (\rho)$ of Eqs. (\ref{Euler1})
demonstrate an inhomogeneity of $M$ associated with the 
Skyrmion cores, Fig.~\ref{conf1}(a).
Detailed analysis of the solutions for
isolated skyrmions has revealed a number
of remarkable phenomena imposed by interplay 
between  the angular ($\theta$)
and modulus ($M$) order parameters
\cite{confinement10}. 

(i) \textit{Collapse of Skyrmions at high fields.}
The solutions of Eqs.~(\ref{Euler1}) exist only
below a critical \textit{collapse} line (Fig. \ref{conf1}).
As the applied field approaches this line the magnetization
in the Skyrmion center $M(0)$ gradually shrinks, and
as $M(0)$ becomes zero the Skyrmion collapses.
This is in contrast with low-temperature
Skyrmions, which exist without collapse even at very large
magnetic fields \cite{JMMM94} because the stiffness
of the magnetization modulus maintains topological stability of Skyrmions.
At high temperatures the softening of the magnetization
allows the order parameter to pass through zero 
and to unwind the Skyrmion core.

(ii) \textit{Crossover of Skyrmion-Skyrmion interactions.}
The coupling of Skyrmions is repulsive in a broad
temperature range, but it starts to oscillate
at high temperatures,  Fig. \ref{conf1} (b).
The equation to describe this crossover in 
the inter-Skyrmion coupling is
\begin{eqnarray}
\label{crossover}
\eta H^* = \sqrt{2 \pm \sqrt{3+4 a}} \;( a + 1 \pm \sqrt{3+4 a}/2 ),
\quad 
T_L =  T_N - 4 T_D, 
\quad
T_D = D^2/(8 A J).
\end{eqnarray}
For $a < -0.5$ this yields a crossover line in the 
phase diagram, Fig.~\ref{conf1}(b).
This is ended in turning point (-0.5, 0).
Another turning point (-0.75, $\sqrt{2} \eta/4$)
indicates the lowest temperature where 
Skyrmions can attract each other 
at certain distances,
owing to the oscillatory character of 
their interactions, the \textit{confinement temperature},
$T_L$ (\ref{crossover}).
This results in energetic confinement of Skyrmions,
as they can form clusters to lower their energy.

(iii) \textit{Confinement.}
For $-0.5 < a < 0.25$ 
line (\ref{crossover}) delimits
a small pocket (III) in the
vicinity of the ordering temperature.
Within this region Skyrmions
can exist only as bound states.
%
%
In the \textit{confinement} region
Skyrmionic states drastically differ
from those in the main part of the phase diagram. 
Confined Skyrmion textures
(hexagonal and square half-Skyrmion lattices, 
Fig. \ref{conf1} (c))
arise from the disordered
state through a rare case of an 
\textit{instability-type nucleation transition}, 
and their field-driven transformation 
evolves by distortions of the modulus
profiles $M(\rho)$ while the equilibrium
periods of the lattices do not
change strongly with increasing
applied field (for details
see \cite{confinement10}).
\subsection{Confinement phenomenon and precursor states}
The magnetic phase diagram of isotropic 
helimagnet (Fig. \ref{conf1} (b))
includes two distinct temperature intervals:
(i) in the main part ($0 < T < T_L$) 
the rotation of the local magnetization vector
determines the chiral modulation, while 
the magnetization amplitude remains constant.
(ii) At high temperatures ($T_L < T < T_N$)
spatial variation of the magnetization modulus
becomes a sizeable effect, and strong interplay
between $M$ and angular variables is the main
factor in the formation and peculiar behaviour
of chiral modulations in this region.
The \textit{confinement temperature} $T_L$ 
\cite{confinement10} provides the scale 
that delineates the border between these 
two regimes in the phase diagram.
The characteristic temperature $T_L$
is of fundamental importance for chiral magnets.
It is of the same order of magnitude as 
the temperature interval $(T_N- T_c) \propto D^2/A$,
where chiral couplings cause inhomogeneous 
precursor states around the magnetic order temperature.
Due to the relativistic origin 
and corresponding weakness of the DM exchange
the shift $\Delta T =T_N - T_L$ is small 
(for MnSi this is estimated to be about 2 K \cite{confinement10}).

Since discovery of chiral modulations
in MnSi family of B20 compounds 
numerous magnetic anomalies have been observed 
near the ordering temperature 
of these helimagnets \cite{Lebech89}.
However, the nature of the chiral 
spin textures in this region of the phase diagram 
is largely unresolved in experiment.
During last years these \textit{precursor} effects
have become a subject of intensive investigations
\cite{Lamago06,Muhlbauer09}
motivated by the expectation to identify Skyrmionic states \cite{Nature06}.

The rigorous solutions for helical and Skyrmionic
modulations recently derived within 
the basic model $w = w_0 +f(M)$ (\ref{density})
and the theoretical description
of novel phenomena attributed
to this region \cite{confinement10} 
give a first explanation of these anomalies.
The stabilization of Skyrmionic textures and 
other chiral modulations near magnetic ordering 
involve the confinement of localized state, but also 
a strong influence of minor energy contributions. 
E.g., the phase diagram of cubic
helimagnets contains defected half-Skyrmion
lattices at small applied fields and densely 
packed full-Skyrmion lattices. 
However, the conformation of these mesophases
and their relative stability will depend on various
small additional effects, such as dipolar couplings,
thermal fluctuations, quenched defects etc.
It would be naive to expect these phase diagrams to 
be simple and to be determined by one dominating mechanism
able to stabilize Skyrmions over helicoids.
Owing to the hierarchy of interactions, the properties and 
stability of the Skyrmion cores will always be provided 
by exchange and Dzyaloshinskii-Moriya couplings via 
the double-twist mechanism. 
But the mesophase formation will be ruled by much
weaker couplings, owing to the localized and frustrated
nonlinear character of these solitonic entities.

\section{Topogical solitons, vortices, Skyrmions...}

\subsection{How the chiral Skyrmion got its name}

Localized solutions of Eq.~(\ref{eq}) have been initially introduced
under name \textit{magnetic vortices} \cite{JETP89} because, to a certain
extent, they are similar to 2D topogical defects investigated 
in magnetism and known as "two-dimensional topological solitons" or "vortices" 
(e.g.  well-known Belavin-Polyakov solutions 
for magnetic vortices \cite{Belavin75}).
On the other hand, the term \textit{skyrmion}
has been conceived in a field rather distant from condensed-matter physics 
and  initially was related to the localized solutions derived by 
Skyrme within his model for low-energy dynamics of mesons and baryons \cite{Skyrme61}.
In fact, the Skyrme model \cite{Skyrme61}
comprises three spatial dimensions,
and the name ``baby skyrmion'' was used
by some field theorists to distinguish
two-dimensional localized states from ``mature''
three-dimensional solutions in the original Skyrme model \cite{Skyrme61}, 
both types of them being topological static solitons.
During the last decades the "skyrmion" has
progressively won currency in general physics
to designate \textit{any} non-singular localized 
and topologically stable field configuration.
Complying with this trend, in 2002 we have renamed
"chiral magnetic vortices" into "chiral skyrmions" \cite{PRB02}.
Ironically, the fate of the localized states 
near ordering transition, as they can decay
by longitudinal magnetization processes, 
betrays that these ``chiral Skyrmions'' 
are not topologically stable, at high enough temperature.

\subsection{'What exactly is a chiral skyrmion'?}

Thus, the term \textit{skyrmion} 
is an umbrella title for smooth localized 
structures to distinguish them from 
singular localized states,
e.g., disclinations in liquid crystal textures \cite{DeGennes}. 
This convention provides only 
a formal \textit{label} for a large
variety of very different solitonic states
from many fields of physics \cite{skyrmion10}.
With respect to the subject of this paper
a "skyrmion"  designates well-defined solutions
of Eq.~(\ref{eq}) which are
\textbf{ (i) localized}, \textbf{(ii) axisymmetric},
and have \textbf{ (iii)  fixed rotation sense}.
Examples of chiral skyrmions have been presented
in Figs. \ref{Skyrmion}, \ref{uniaxial0}.
The axisymmetric structure of
the skyrmion core and its localized character are retained
in bound states as skyrmion lattices \cite{Nature06,JPC10}.
This reflects the particle-like character
of chiral skyrmions and the most general 
features of their energetics (Fig. \ref{uniaxial0} (a),
and Refs. \cite{Nature06,JPC10}).
Skyrmions as countable entities can be arranged in 
various ways to create dense magnetic textures.
This is the essence of Skyrmionic matter and entails
the possibility to form a variety of mesophases 
with crystalline, but also with liquid-like or 
glassy large-scale structure in chiral magnetism.

Alternative approaches construct skyrmionic 
textures from crossing plane waves (helices) as fundamental
modes of so-called spin-spiral crystals \cite{Muhlbauer09,Binz06}.
In \cite{Muhlbauer09}  a skyrmion lattice is composed
of three helices superimposed under an angle of 120 degree. 
According to \cite{Muhlbauer09} such a "triple-Q antiskyrmion lattice"
ansatz even reaches the global free energy minimum in the A phase of MnSi.
This construction is inconsistent with the properties of Skyrmions.  
Chiral skyrmionic textures incorporate isolated or embedded axisymmetric lines.  
The notions of a "spin-spiral crystal" \cite{Binz06} or
"multi-Q skyrmions"  \cite{Muhlbauer09}
are misconceptions because they blend 
mutually exclusive ideas of particle-like localized 
skyrmions from one side and delocalized plane waves from the other.
The radial structure of the chiral Skyrmion cannot 
be reduced to a superposition of harmonic helical waves. 
Calculations based on such multi-Q ansaetze 
predict incorrect phase diagrams.
E.g., the hexagonal Skyrmion lattice is stable in remanent state
at $H=0$ \cite{JMMM94}, but in \cite{Muhlbauer09} an instability at a finite
field is presented for the variational hexagonal spin-spiral solution.
The idea of a spin-spiral crystal 
ruled by one harmonic mode, where contributions of 
higher-order modes are small in some sense \cite{Muhlbauer09,Binz06}, 
also is at variance with the nucleation type of transition 
actually observed in the chiral Skyrmion textures. 
This idea wrongly places the Dzyaloshinskii theory 
in a class of models for modulated states 
with instability type of transitions.

\section{Conclusions: What makes  chiral skyrmions interesting?}

In non-centrosymmetric magnets 
chiral magnetic skyrmions arise as a result of 
the specific stabilization mechanism imposed by the
handedness of the underlying crystal structure
\cite{Dz64,JETP89,Nature06}. 
In centrosymmetric
magnets such solutions are radially unstable and
collapse spontaneously under the influence
of the applied magnetic field or intrinsic 
short-range interactions.
In nonlinear field models skyrmion states 
can be stabilized by higher order spatial derivatives
(often this is refered to \textit{Skyrme mechanism}). 
In condensed-matter systems 
there are no physical interactions providing such 
energy contributions. \textit{Chiral interactions}
present a unique mechanism to stabilize skyrmion 
states in ordered condensed-matter systems.
This singles out chiral systems (including non-centrosymmetric
magnets, multiferroics, liquid crystals, and metallic nanostructures
with induced chiral interactions) into a particular class of materials
with skyrmionic states.

\textit{Acknowledgment.} 
We thank 
S. Bl\"{u}gel, S.V. Grigoriev, G. G\"{u}ntherodt, R. Sch\"{a}fer, W. Selke, 
and  H. Wilhelm, for discussions.  
Support by DFG project RO 2238/9-1 is gratefully acknowledged.

%
%
%
%
%
%

\end{document}